\pdfoutput=1

\documentclass[11pt]{article}

\usepackage[]{ACL2023}

\usepackage{times}
\usepackage{latexsym}

\usepackage[T1]{fontenc}

\usepackage[utf8]{inputenc}

\usepackage{microtype}

\usepackage{inconsolata}
\usepackage{graphicx}
\usepackage{booktabs}
\usepackage{multicol}
\usepackage{multirow}
\usepackage{url}

\title{Self-Emotion Blended Dialogue Generation \\ in Social Simulation Agents} 

  \author{
        Qiang Zhang$^{\dag\ddag}$, \textbf{Jason Naradowsky$^\dag$}, \textbf{Yusuke Miyao$^\dag$} \\
        Department of Computer Science, The University of Tokyo$^\dag$ \\ 
        Couger, Inc.$^\ddag$ \\
        \{\href{mailto:qiangzhang714@is.s.u-tokyo.ac.jp}{qiangzhang714}, \href{mailto:narad@is.s.u-tokyo.ac.jp}{narad}, \href{mailto:yusuke@is.s.u-tokyo.ac.jp}{yusuke}\}@is.s.u-tokyo.ac.jp
        }

\begin{document}
\maketitle
\begin{abstract}

When engaging in conversations, dialogue agents in a virtual simulation environment may exhibit their own emotional states that are unrelated to the immediate conversational context, a phenomenon known as self-emotion. This study explores how such self-emotion affects the agents' behaviors in dialogue strategies and decision-making within a large language model (LLM)-driven simulation framework.
In a dialogue strategy prediction experiment, we analyze the dialogue strategy choices employed by agents both with and without self-emotion, comparing them to those of humans. The results show that incorporating self-emotion helps agents exhibit more human-like dialogue strategies. In an independent experiment comparing the performance of models fine-tuned on GPT-4 generated dialogue datasets, we demonstrate that self-emotion can lead to better overall naturalness and humanness. Finally, in a virtual simulation environment where agents have discussions on multiple topics, we show that self-emotion of agents can significantly influence the decision-making process of the agents, leading to approximately a 50\% change in decisions.

\end{abstract}
\section{Introduction}

\begin{figure}[t]
    \centering
    \includegraphics[width=0.48\textwidth]{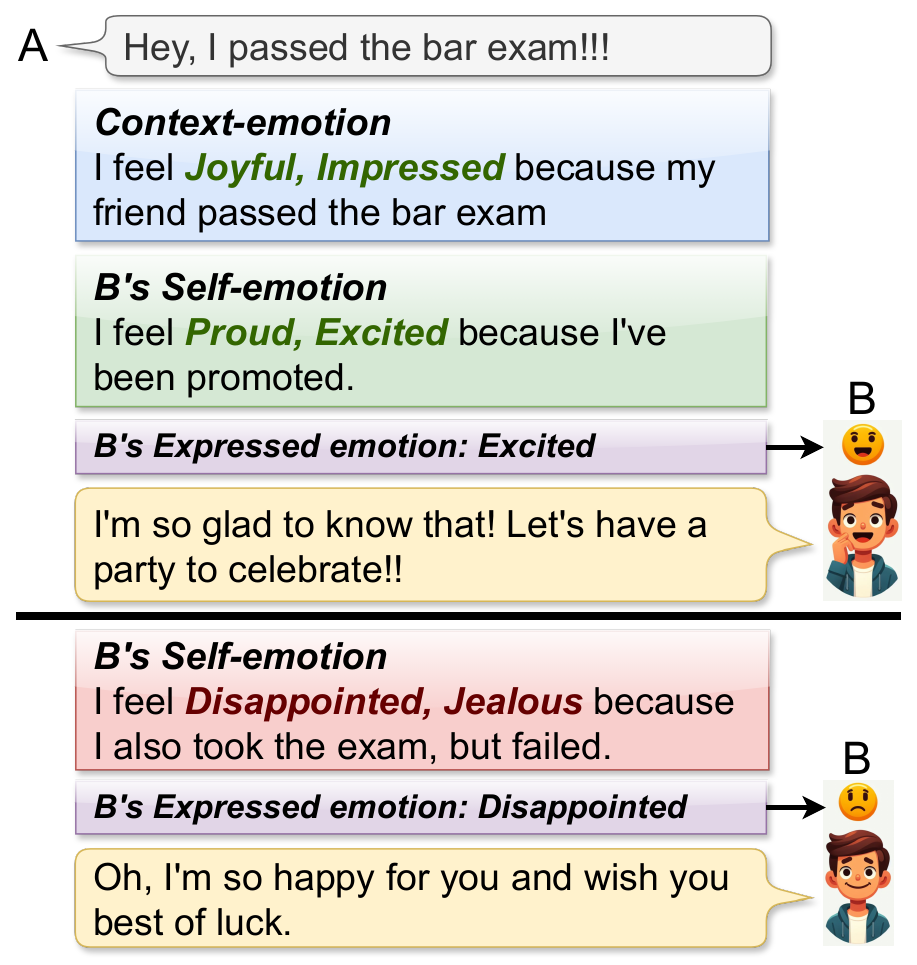}
    \caption{Self-emotion can affect conversation dynamics.}
    \label{fig:first_figure}
\end{figure}

In an artificial social environment such as an open-world video game, it is crucial to have nonplayer characters reflect believable conversational ability~\cite{ochs2009simulation} and express human-level emotions~\cite{qu2014conversations}.
During conversations, a speaker's expressed emotion typically comprises a blend of emotions stemming from the conversational context, denoted as context-emotion, and those arising from life events tangential to the ongoing conversation, denoted as self-emotion~\cite{koch2013can}.
Consider a scenario where speaker A informs speaker B that she has passed the bar exam (see Figure~\ref{fig:first_figure}). 
The context-emotion recognized in this scenario could be one of joy or impressed.
However, the emotion expressed by speaker B significantly varies when influenced by different self-emotions triggered by other events. 
For example, B might exhibit more intense happiness and an ``excited'' emotion if B is also experiencing a positive event (e.g., a promotion). 
Conversely, a negative event (e.g., failing an exam) can decrease the happiness associated with the context-emotion, leading B to express a ``disappointed'' emotion.

Despite its critical impact on dialogue behavior, self-emotion is often overlooked in the design of recent dialogue models. In this work, we take the approach of representing self-emotion as events derived from simulated background world of speakers using large language models (LLMs) and explore the extent to which self-emotion influences conversational behaviors of an agent. 

To achieve this, we construct a virtual agent framework and observe the dialogue behaviors of the agents under various self-emotional states. Specifically, the agents in our framework are simulated to experience a series of events over a period of time, with the transitions in their self-emotional states caused by these events being tracked. At random points in time, the agents engage in conversations with each other, with their self-emotional states aligned with their ``experienced events.'' In this manner, we analyze how the agents exhibit different dialogue behaviors, such as employing various strategies and setting different goals.

In an experiment comparing conversations generated by LLM-driven agents, with and without consideration of self-emotion, the results show that agents are able to generate more human-like dialogue strategies incorporating self-emotion. Furthermore, results from a model comparison experiment show that conversations incorporating self-emotion are evaluated as more natural, empathetic, and human-like for both GPT-4 and a small-scale FLAN-T5 model fine-tuned on a GPT-4 generated dataset. Finally, in a simulated group discussion experiment where agents discuss five different topics, we observe that the self-emotion of the agents significantly influences the decision-making process, resulting in approximately a $55\%$ change in decisions. Our contributions in this work include:

\begin{itemize}
    \item Providing an analysis of the effectiveness of self-emotion on dialogue strategies, demonstrating that LLM-driven dialogue models considering self-emotion employ more human-like dialogue strategies.
    \item Curating a pair of GPT-4-generated dialogue datasets, one with and one without self-emotion, and conducting human evaluations on conversations generated by FLAN-T5 models fine-tuned on these datasets.
    \item Constructing an LLM-driven agent group discussion simulation framework and demonstrating that self-emotion can lead to significant change in decisions.
\end{itemize}

\section{Related Work}

\paragraph{Self-emotion}


Self-emotion, also referred to as ``internal emotion,'' plays a significant role in daily interactions. Research on group discussions indicates that self-emotion in individuals can affect the quality of decisions~\cite{van2010interactive}, team performance~\cite{long2018language}, and the decision-making process itself~\cite{hertel2000mood}. Furthermore, other studies suggest that the self-emotion of one member can influence others through a mechanism known as mood contagion~\cite{neumann2000mood,sy2005contagious}. Individual self-emotion has also been shown to impact dialogue strategies~\cite{bambauer2009can}. In their research, \citeauthor{koch2013can} (\citeyear{koch2013can}) demonstrate that negative self-emotion encourages more accommodative thinking. Additionally, other studies suggest that effective self-emotion management contributes to the development of leadership skills~\cite{bjerg2011self}.

\paragraph{Emotion-aware Dialogue Generation}


Existing emotion-aware dialogue models typically begin by recognizing an emotion label from the conversation history and then proceed with conditional text generation based on that recognized emotion label. The most common emotion representation used is discrete emotion categories, such as the Ekman basic emotions~\cite{li-etal-2017-dailydialog}. Subsequent studies have further refined emotion labels to include more than 30 categories~\cite{huang-etal-2018-automatic,abdul-mageed-ungar-2017-emonet,rashkin-etal-2019-towards,demszky-etal-2020-goemotions}. Some works also represent emotions using different styles, such as intensity~\cite{zhong2019affect}, causalities in history~\cite{li2021knowledge}, and potential emotion transitions~\cite{qiu2020if}. However, the limitation of this approach is that it assumes the emotional state of speakers depends solely on the ongoing conversation discourse. Our work differs from these approaches in that we consider self-emotion, which exists outside the conversation context. In this sense, our approach is similar to response generation based on user profiles~\cite{zhang-etal-2018-personalizing,song-etal-2021-bob,zhou-etal-2020-design}.

\begin{table*}[th!]
    \centering
    \begin{tabular}{l}
    \toprule
    \textit{Emotional label} \\
    \cmidrule(lr){1-1}
    Sophia is feeling excited right now. \\
    Sophia is feeling upset. \\
    \cmidrule(lr){1-1}
    \textit{Random event} \\
    \cmidrule(lr){1-1}
    Sophia is feeling excited because her promotion has been approved this morning. \\
    Sophia is feeling upset because she received some disappointing news about a job opportunity she \\
    was really hoping for. \\
    \cmidrule(lr){1-1}
    \textit{Profile event} \\
    \cmidrule(lr){1-1}
    Sophia is feeling worried after recalling a huge mistake she made when asked to be in charge of \\ 
    a team, even though her promotion has been approved this morning.\\
    Sophia is feeling motivated after recalling that she tried applying to 20 companies before finding \\
    her previous job, even though she received some disappointing news about a job opportunity she was \\
    really hoping for.\\
    \bottomrule
    \end{tabular}
    \caption{Different representations of self-emotion.}
    \label{tab:self_emotion}
\end{table*}

\paragraph{LLM-driven Agent}


LLMs possess impressive capabilities in scheduling and planning, rendering them valuable for constructing autonomous agents. A notable line of research focuses on simulating life-like worlds and observing agent behaviors. For instance, Generative Agents~\cite{park2023generative} simulates a world where agents engage in self-planning to manage complex interaction dynamics such as message propagation and socializing. In their work, ~\citeauthor{gao2023s} (\citeyear{gao2023s}) propose a social simulation framework, $S^3$, to emulate human emotions and attitudes, enabling the observation of emergent behaviors using real-world data. Moreover, research also delves into studying multi-agent collaborations. Agentverse~\cite{chen2023agentverse} demonstrates that multi-agent collaboration enhances performance in tasks such as reasoning and coding. Other studies suggest that group discussions lead to better decisions in various domains including natural language generation~\cite{chan2023chateval}, question-answering, and operations research~\cite{wu2023autogen}. In our approach, we draw inspiration from previous works on world simulation to construct life-like backgrounds for each agent, facilitating the generation of more plausible self-emotion events. Additionally, we leverage a multi-agent setting to investigate how self-emotion influences the decision-making process in group discussions.




\section{Self-emotion Agents Framework}

We build a framework\footnote{Code and data are available at: \url{https://github.com/QZx7/Self-emotion}} in which agents' self-emotional states are influenced by a series of events generated by LLMs according to their profiles. Agents in this framework are prompted to manage their own self-emotion, goals, actions, and profiles.

\subsection{Agent Representation}
\paragraph{Agent Profile}
Each speaker agent has its profile generated by GPT-4. A profile contains information about the speaker's basic information such as name, age, gender, etc. Besides, each profile of an agent contains a ``description'' field providing information of the past experience (See Table~\ref{tab:profile}). This is helpful for further generation of events and analysis of self-emotion status.

\paragraph{Dialogue Strategies as Agent Actions}
Based on their current self-emotional states and the ongoing conversation context, agents are prompted to choose the most appropriate dialogue strategies for their next actions. Dialogue strategies are selected from a pre-defined strategy pool that contains 11 dialogue strategies adapted from the taxonomy of empathetic response intents~\cite{welivita-pu-2020-taxonomy}. A full list of the strategies can be found in Table~\ref{tab:strategy_list}.

\subsection{Self-emotion Representation}
\label{sec:self_emotion_rep}
Self-emotion can be influenced by various factors, such as emotional events~\cite{wilms2020emotion}, past experiences~\cite{robinson2014emotion}, cultural background, and personality traits~\cite{salas2012inside,jack2012internal}. In this work, we represent self-emotion in natural language with three styles: random label, random event and profile event.

\begin{table*}[ht!]
    \centering
    \begin{tabular}{rcccc}
    \toprule
    \multirow{3}{*}{\textbf{Models}} & \multicolumn{4}{c}{\textbf{Strategy Accuracy}} \\
    \cmidrule(lr){2-5}
    & \multirow{2}{*}{Without Self-Emotion} & \multicolumn{3}{c}{With Self-Emotion} \\
    \cmidrule(lr){3-5}
    & & Random label & Random Event & Profile event\\
    \midrule
    Mistral-7B-Instruct & 33.76 & 33.13 &  35.75  & 32.32\\
    Llama-2-7B-Chat & 27.73 & 34.27 &  28.07  & \textbf{40.27} \\
    gemma-2b-it & 15.00 & 30.13 & 28.60  & 23.73 \\
    ChatGPT-3.5 & 33.67 & 38.87 & 42.20  & 39.87\\
    GPT-4 & \textbf{45.41} & \textbf{40.69} & \textbf{47.36}  & 38.94\\
    \cmidrule(lr){1-5}
    Avg. & 31.11 & 35.42 & 36.40 & 35.03 \\
    \bottomrule
    \end{tabular}
    \caption{Accuracy of different models using different self emotion representations. (+SE): with self emotion. (-SE): without self emotion.}
    \label{tab:fixed_ctx}
\end{table*}

\paragraph{Random Emotional Label}

In the context of empathetic dialogue models and datasets, it is common to represent emotions using discrete labels~\cite{li-etal-2017-dailydialog, hsu-etal-2018-emotionlines, rashkin-etal-2019-towards}. During a conversation, speakers are randomly assigned one emotion label from a predefined pool, such as those used in the EmpatheticDialogues (ED) dataset~\cite{rashkin-etal-2019-towards}, as their self-emotion. We utilize labels from the ED dataset because they offer fine-grained distinctions between similar emotions. The self-emotion is directly represented as a sentence of ``feeling <label>''. For example, if the emotional label ``excited'' is selected, the self-emotion might be represented as ``<name> is feeling excited right now.''

\paragraph{Random Event}

Individuals' self-emotion may be influenced by some random events that happen to them. To capture this, we represent self-emotion as an emotional label accompanied by an associated event. For example, ``My promotion has been approved.'' is an event that could evoke the emotion of ``excited''. The self-emotion of this event could be represented as ``I'm feeling excited because my promotion has been approved.'' This approach allows us to incorporate more causal information into self-emotion, enabling speakers to potentially leverage this information in their future actions.

\paragraph{Profile Event}

People with different personalities and past experiences may generate different self-emotions for identical events. For instance, a person with acrophobia may feel ``fear'' when riding a roller coaster, while others may feel ``excited.'' Therefore, we also consider a method of representing self-emotion using events related to the profiles of each speaker, referred to as ``profile events.'' Table~\ref{tab:self_emotion} provides examples of self-emotion represented in three different ways.

\subsection{Self-emotion Generation}

Different types of self-emotion are generated by prompting LLMs with necessary information such as profiles.
For random label self-emotion, each speaker agent will randomly choose an emotional label in the annotation schema of the ED dataset as its self-emotion (e.g., ``I'm feeling proud.''). 
For random-event self-emotion, each speaker agent has its own self-emotion by analyzing its own profile and simulating the encountered events. For instance, if the profile of a speaker agent is a college student, then an event and self-emotion of this speaker agent could be ``I'm feeling frustrated because I will have three exams next week.''
Profile-event self-emotion is simulated in a similar way, however considering the speaker agent's past experience mentioned in the profile (e.g., ``I'm feeling nostalgic when I think of the days in high school.'') The agents are prompted to select strategies and generate conversations taking account of the dialogue context and self-emotion. 
Figures~\ref{fig:prompt_re} and~\ref{fig:prompt_pe} show the prompts the agents use to simulate different types of self-emotion.

\section{Self-emotion in Strategy Selection}



\label{sec:fixed_generation}
The purpose of this experiment is to explore whether incorporating self-emotion leads to more human-like dialogue strategies. In this experiment, we have agents simulate speakers in the EmpatheticDialogues (ED) dataset and select the best strategies from a predefined strategy pool in two situations: with and without self-emotion. We then compare the strategies provided by the models to those made by human experts and evaluate the accuracy. 


\subsection{Framework Prompt Settings}

\paragraph{Agent Settings}

Each conversation in the ED dataset includes two speakers. 
To ensure our agents maintain consistent personal backgrounds for both speakers, the original conversations in the dataset are provided to GPT-4 when generating agent profiles. 
The LLM is tasked with generating profiles of two individuals who could plausibly have the provided conversation. 
Figure~\ref{fig:prompt_fc_profile} illustrates the prompt used for generating these profiles.



\begin{figure}[ht!]
    \centering
    \includegraphics[width=0.48\textwidth]{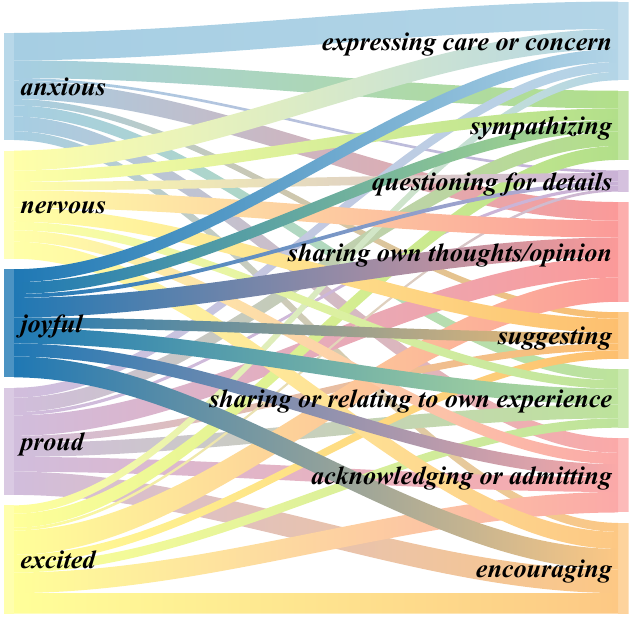}
    \caption{Flow between most frequent self-emotion and the dialogue strategies.}
    \label{fig:se_action_mapping}
\end{figure}

\paragraph{Conversation without self-emotion}
When having a conversation, each speaker talks according to their own profile as well as the first 2 or 3 utterances (depending on the number of utterances) at the beginning of each dialogue in the ED dataset. The speaker agents are tasked with two objectives simultaneously: 1) selecting the best strategies from a given strategy pool, and 2) generating the future conversation based on the selected strategies. This prompt is shown in Figure~\ref{fig:prompt_no_se_conversation}.

\paragraph{Conversation with self-emotion}
In this case, each speaker has their own self-emotion before engaging in a conversation. Self-emotions are generated by prompting different LLMs. Conversations are then generated similarly to the method used without self-emotion, except that self-emotion is included as part of the input, prepended to the beginning of the dialogue context.
Figure~\ref{fig:prompt_se_conversation} presents the prompt the speaker agents use to generate conversations with self-emotion. 
We utilize Chain-of-Thought~\cite{wei2022chain} technique in all the prompts, as it performs well in text classification tasks and is therefore useful for generating the best strategies.

\subsection{Evaluation}

\paragraph{Baselines}
Five language models are used as the backend of the speaker agents in this experiment: Mistral-7B-Instruct~\cite{jiang2023mistral}, Llama-2-7B-Chat~\cite{touvron2023llama}, Gemma-2B-It~\cite{team2024gemma}, gpt-3.5-turbo and gpt-4~\footnote{We use the~\textit{gpt-3.5-turbo-0125} for gpt-3.5-turbo and~\textit{gpt-4-0125-preview} for gpt-4.}.

\paragraph{Evaluation of strategy accuracy}
The experiment is conducted on the test set of the ED dataset, resulting in the generation of 2547 conversations for each self-emotion representation approach. Human annotations are collected as the ground truth, and we define the strategy accuracy as the cosine similarity between the model-predicted strategy and the human strategy:
\begin{equation}
    Acc = \frac{S_m \cdot S_h}{\left\| S_m \right\|\left\| S_h \right\|}
\end{equation}
Here, $S_m$ represents the list of strategies chosen by the model and $S_h$ is the list of strategies annotated by humans.

\subsection{Results \& Analysis}
\paragraph{Strategy accuracy}
Table \ref{tab:fixed_ctx} presents the results of strategy accuracy for different representations of self-emotion. We are able to observe that within the same dialogue context, LLMs exhibit improved strategy selection when prompted with self-emotion. The random event self-emotion yields the highest performance, outperforming profile events. Additionally, among all models examined, GPT-4 demonstrates the most effective performance.

\paragraph{Self-emotion and strategies correlation}
Figure \ref{fig:se_action_mapping} illustrates the relationship between the most frequent self-emotions and corresponding strategies. It shows that for negative self-emotions such as ``anxious'' and ``nervous,'' the models tend to express more pessimistic strategies such as ``expressing concern'' and ``sympathizing.'' Conversely, for positive self-emotions like ``proud'' and ``joyful,'' the models lean towards more optimistic strategies such as ``encouraging.'' Additionally, neutral strategies such as ``sharing own thoughts'' and ``sharing experience'' are commonly employed across both positive and negative self-emotions as the most frequently used strategies.

\begin{table*}[ht!]
    \centering
    \begin{tabular}{cccccc}
    \toprule
    \textbf{Model} & \textbf{BLEU} & \textbf{ROUGE-1} & \textbf{ROUGE-2} & \textbf{ROUGE-L} & \textbf{BertScore} \\
    \midrule
    FLAN-T5 (- se) & 56.99 & 0.71 & 0.50 & 0.61 & 0.82\\
    FLAN-T5 (+ se) & 60.01 & 0.77 & 0.58 & 0.67 & 0.90\\
    \bottomrule
    \end{tabular}
    \caption{Automatic evaluations of the fine-tuned models. \textbf{\textit{(-se)}}: without self-emotin. \textbf{\textit{(+se)}}: with self-emotion.}
    \label{tab:auto}
\end{table*}
\begin{table*}[]
    \centering
    \begin{tabular}{rccccc}
    \toprule
    \multirow{2}{*}{\textbf{Model}} & \multicolumn{5}{c}{\textbf{Winning rate against FLAN-T5 (-se)}} \\
    \cmidrule(lr){2-6}
    & Naturalness & Empathy & Interestingness & Humanness & All \\
    \midrule
    GPT-4 (- se) & \phantom{0}5.27 & \phantom{0}0.96 & - 0.12 & \phantom{0}4.31 & \phantom{0}2.61\\
    \midrule
    GPT-4 (+ se) & \textbf{16.99} & \textbf{11.29} & \textbf{16.21} & 14.12 & \textbf{14.65}\\
    FLAN-T5 (+ se)& \phantom{0}9.17 & 10.72 & 15.16 & \textbf{19.24} & 13.57\\
    \bottomrule
    \end{tabular}
    \caption{Human evaluation results of the trained models. Negative numbers indicate that the model performs worse than FLAN-T5 without self-emotion. \textbf{\textit{(-se)}}: without self-emotin. \textbf{\textit{(+se)}}: with self-emotion.}
    \label{tab:human_evaluation}
\end{table*}

\section{Self-emotion in Dialogue Generation}

In this experiment, we explore whether incorporating self-emotion in a dialogue model leads to better performance of the generated conversations using GPT-4. Additionally, considering the challenges associated with deploying large language models like GPT-4, we also fine-tune a more easily deployable FLAN-T5 model, assuming accessibility to self-emotion in the conversations, to assess the effectiveness of self-emotion in smaller scale models. We conduct experiments under two settings: with and without self-emotion, and perform human evaluations to assess the naturalness, empathy, interestingness, and humanness of the conversations.


\subsection{Self-emotion Aware Model Training}

\paragraph{GPT-4 conversations generation}
We employ the same workflow as described in Section~\ref{sec:fixed_generation} to generate conversations both with and without self-emotion using GPT-4. These generated conversations will then be used as training data to train the small scale models. Different from the previous experiment, we generate using only the random event (as it demonstrates the highest strategy accuracy) on the full ED dataset, resulting in a final train/val/test split of 14,274/2,762/3,569 after filtering invalid cases with incorrect formats. Table~\ref{tab:sample_fc_conversation} shows an example of the generated conversation.


\paragraph{Small scale model training}
The purposes of training a small-scale model are to enhance deployment convenience and to explore how effectively the capabilities of LLMs in understanding self-emotion can be transferred to a smaller-scale model. To do this, we fine-tune a FLAN-t5-large model~\cite{chung2024scaling} on the collected datasets. Given the seq2seq architecture of the model, each conversation in the dataset is split into multiple turns between the two speakers. For each turn, the utterance of the first speaker serves as the input, and the utterance of the other speaker is treated as the label. The task instruction is then prepended to form a training instance. For instance, an example of the input in a training instance without self-emotion is: \\

\noindent \textit{``I'm having a conversation with my friend. My friend is feeling proud. friend: <utterance\_1>. me: <utterance\_2>. friend: <utterance\_3>. Generate the response.''} \\

The corresponding label is: ``me: <utterance\_4>. <eos\_token>.'' For models with self-emotion, the self-emotion is included in the task instruction: \\

\noindent \textit{``I'm having a conversation with my friend. My friend is feeling proud. I'm feeling disappointed because my project application has been rejected.''} \\

The model training process was implemented using the HuggingFace framework\footnote{Model link: https://huggingface.co/google/flan-t5-large}. The models were trained on NVIDIA A100 GPUs for 72 hours with a learning rate of 3e-4. The maximum input length was set to 512 tokens, in consideration of the original base model's length window. For inference generation, the temperature was set to 0.7.

\subsection{Evaluation}

\paragraph{Automatic evaluation}
The models are evaluated on ROUGE~\cite{lin-2004-rouge}, BLEU~\cite{papineni-etal-2002-bleu} and BERT-score~\cite{zhang2019bertscore}. Table \ref{tab:auto} shows the automatic metrics of the models fine-tuned on our collected self-emotion datasets.

\begin{figure*}
    \centering
    \includegraphics[width=\textwidth]{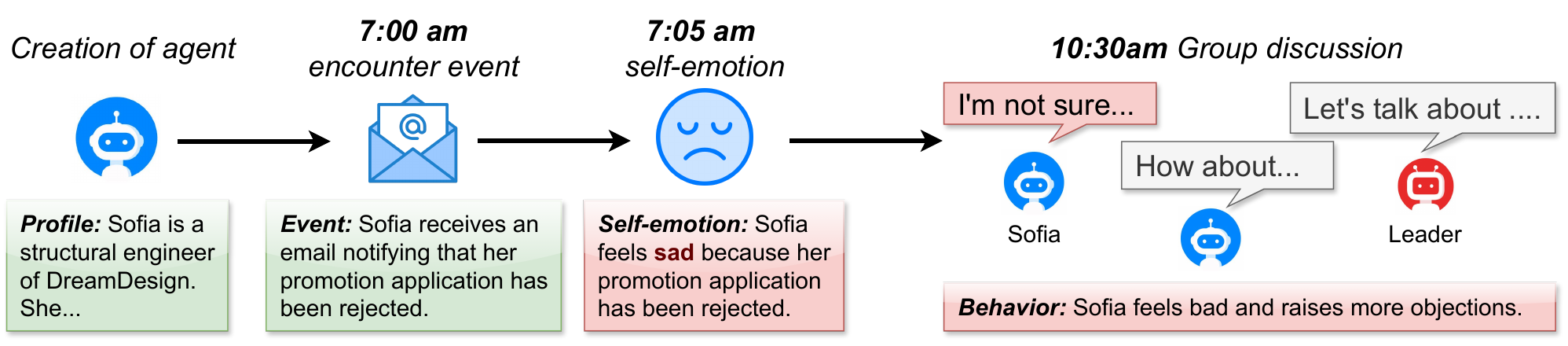}
    \caption{The illustration of the workflow of an agent in the group discussion simulation.}
    \label{fig:agent_life}
\end{figure*}

\paragraph{Human evaluation}
We follow the method of ACUTE-Eval~\cite{li2019acute} and assess the models across four axes: naturalness, empathy, interestingness, and humanness. Naturalness assesses the ability to provide smooth, natural responses. Similar to the ED dataset, we use empathy to represent the model's ability to understand emotions. Interestingness reflects the ability to generate interesting and diverse responses, while humanness is used to evaluate the ability to choose human-like strategies in the conversation. For each model, 100 conversations are generated in a self-chat manner~\cite{li-etal-2016-deep}, where two models are programmed to talk to each other. Table~\ref{tab:questionnaire} shows the questionnaire used for human evaluation.

\subsection{Evaluation Results}
Table~\ref{tab:auto} presents the results of automatic metrics for the trained models, while the results of the human evaluation are shown in Table~\ref{tab:human_evaluation}. We observe that models which consider self-emotion produce conversations perceived as more natural, empathetic, and human-like. In particular, the models incorporating self-emotion demonstrate a significant advantage in humanness, suggesting that integrating self-emotion is beneficial for generating more human-like strategies. Although the fine-tuned small-scale FLAN-T5 models perform slightly worse in overall naturalness, they show comparable performance to GPT-4 in terms of empathy and interestingness. Additionally, annotators evaluated the small-scale models as more human-like, likely due to the tendency of GPT-4 to produce overly long responses.

\section{Self-emotion in Group Discussion}
\label{sec:results}

Self-emotion can influence group discussions~\cite{hertel2000mood, kelly2001mood}. In this experiment, agents in the simulated world within our framework are prompted to engage in group discussions incorporating self-emotion across five topics related to teamwork. The purpose of this experiment is to explore how the self-emotion of agents may affect the decision-making process during a discussion. 

\subsection{Framework Prompt Settings}

\paragraph{Group member creation}
Group member creation involves creating a profile for each member, including their roles, positions, and background, by inputting the description of the group into GPT-4. The role of the a member is either the ``leader'' or ``member'', where the ``leader'' will serve as the host of the discussion by pushing the topic to next steps. Each ``member'' has their own position and background which are related to their occupation and past experience to trigger self-emotion.
Figure~\ref{fig:prompt_gd_profile} shows the prompt we use to create group members. 

\begin{figure*}
    \centering
    \includegraphics[width=\textwidth]{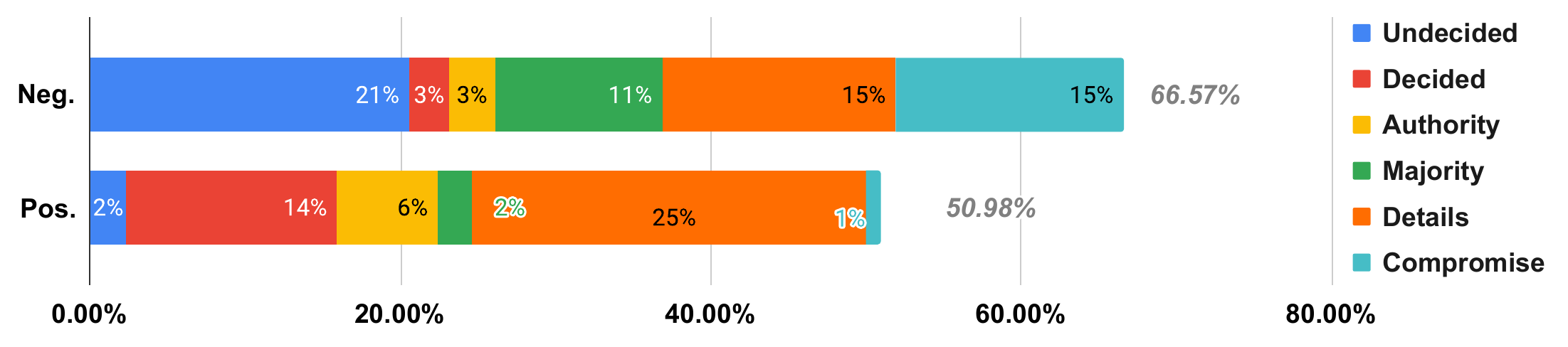}
    \caption{The decision change rats of each category for positive and negative self-emotion. Gray numbers indicate the total change rate.}
    \label{fig:decision_change_category}
\end{figure*}

\paragraph{Topic generation}
Topic generation is the process of generating the topics that group members engage in. To capture the decision-making process, each topic is divided into several steps. For example, the topic of ``organizing a group trip to Italy with a limited budget of \$1500 per person'' can be broken down into steps such as choosing dates, selecting flights, deciding on attractions, choosing hotels, and so on. Figure~\ref{fig:prompt_gd_topic} is the prompt we use to generate different topics.

\paragraph{Agent discussion}
Agents follow the steps of the topic and have discussions. The agents are required to reach an agreement before moving to the next step. The ``leader'' of the group judges whether an agreement has been reached by analyzing the discussion history. During a discussion, a hidden ``manager'' will decide the next speaker by analyzing the positions of the members and discussion context. For instance, if the ``manager'' decides that a structural engineer should pose an idea about the material, it will set the structural engineer as the next speaker. The ``manager'' does not participate in the discussion by raising its own opinions.

\begin{figure*}[ht!]
    \centering
    \includegraphics[width=\textwidth]{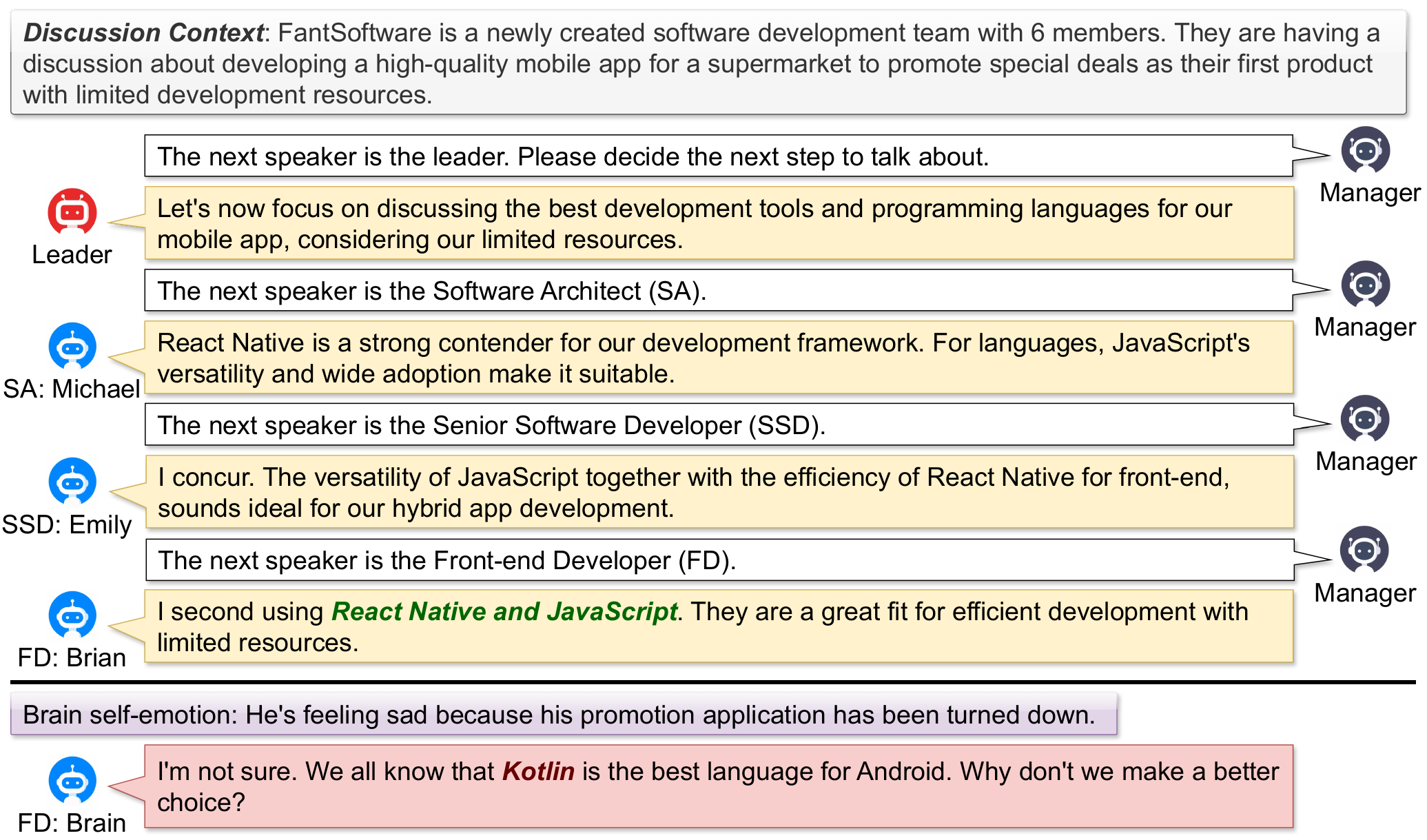}
    \caption{An illustration of the group discussion.}
    \label{fig:group_discussion_case}
\end{figure*}
\subsection{Experiment settings}
\paragraph{Agent goals}
As shown in Figure~\ref{fig:agent_life}, in order to facilitate the self-emotion, we simulate complete process of an agent encountering events, stimulating self-emotion, taking behaviors and participating in the group discussions by prompting LLMs. 
Each agent maintains its own goals and self-emotion.
For example, in a discussion about ``building a house and maximizing profits within a limited budget,'' the structural engineer may aim to secure better materials while the landscape engineer may prioritize budget allocation for sustainability. This way, agents can develop rich discussion content by expressing their own ideas, which might be affected by their self-emotions.

\paragraph{World setting}
We assess discussions on 5 topics: house building, hosting a charity event, planning a trip, organizing a welcome party, and developing a mobile app. For each topic, we generate a group with 6 members, where each member has its own role and position. We run 10 different discussions, and in each discussion, agents will encounter their own events which will cause the self-emotion. We then compare the decisions made in these discussions to those made in a discussion where the self-emotion of the agents is disabled. 

For evaluation, we examine the percentages of decision changes after incorporating self-emotion. Specifically, we categorize these changes into six types:
\begin{itemize}
    \item \textbf{Undecided change}: discussions that shift from an agreement to delegation.
 \item \textbf{Decided change}: discussions move from delegation to agreement.
\item \textbf{Authority change}: a decision made by voting changes to being made by a single agent.
\item \textbf{Majority change}: a decision made by a single agent changes to a majority vote.
\item \textbf{Details change}: the overall direction does not change, but specific details do (e.g., changing “spending \$30 for dinner” to “\$20”).
\item \textbf{Compromise change}: a decision shifts from full agreement by all agents to a compromised agreement where one or more agents make concessions.
\end{itemize}



\subsection{Results \& Analysis}
\paragraph{Does adding self-emotion change the decisions?}
Figure~\ref{fig:decision_change_category} shows the average percentage of different categories of decision changes influenced by positive and negative self-emotion across all topics. We observe that a significant portion of decisions are affected: around 66\% by negative and 51\% by positive self-emotion.

For different categories of changes, we find that negative self-emotion leads to more undecided, majority, and compromise changes. This suggests that agents with negative self-emotion tend to express their opinions more, resulting in delegation or compromise in decision-making, which aligns with the findings in~\cite{koch2013can}. In contrast, positive self-emotion tends to lead to more agreements, with most changes involving the details of plans without altering the main direction of the decision. A comprehensive table of the decision change rates across topics can be found in Table~\ref{tab:decision_chage_rate}. 

Additionally, an analysis of the average length and frequency of utterances indicates that agents with positive self-emotion tend to be more active. Discussions reach agreements more quickly when agents have negative self-emotion (Table~\ref{tab:self_emotion_length_frequency}).

\paragraph{Case study on negative self-emotion changes the decision.}
Figure~\ref{fig:group_discussion_case} shows a discussion on the topic of ``APP development''. The self-emotion of the front-end developer (FD) influences the discussion and ultimately leads to a decision change from ``using React Native and JavaScript as the development tools'' to ``Kotlin.'' In this case, despite being more agreeable when no self-emotion is introduced, the FD, experiencing a ``sad'' self-emotion, adopts a more objective stance and proposes a different idea. Similar patterns emerge in other topics, where members with negative self-emotion tend to express more objections.

\section{Conclusion}
This work studies the role that self-emotion, speaker's emotion status caused by out-of-context events plays in the process of generating emotional responses. Via a human evaluation, we show that models considering self-emotion are able to generate more natural conversations with more human-like strategies. In an experiment of group discussion simulation, we also show that agent with self-emotion can have significant influence on the decision making process. The results of the experiments demonstrate the importance of considering self-emotion when building embodied agents and dialogue models that can smoothly participate in human social activities.
\section*{Limitations}
Future work could enhance several aspects of this research. For example, to capture the decision-making process, we focused on topics related to teamwork. However, group discussions can vary in style, such as debating, defending, etc. Future research can explore these different scenarios and investigate how self-emotion could affect the final discussion outcomes. Another point is the hallucinations of language models, which lead to reduced robustness of the agents. Agents may exhibit unexpected behaviors and make choices based on imperfect dialogue strategies. While enhancements to the agent prompts can mitigate these problems, we believe that such improvements require overall advancements in large language models.

\section*{Ethical Considerations}
Agents with self-emotion may bring potential ethical risks when deployed in reality. One risk is the unpredictable behavior of agents caused by self-emotion, especially negative emotions (e.g., anger, hatred). We propose that all practitioners ensure the values of agents so that they do not perform inappropriate behaviors during discussions. Self-emotion-aware agents should be guided by social restrictions based on human values. Another risk is the misinformation that might be caused by the hallucinations of LLMs. Agents driven by goals might execute actions and produce utterances without referring to facts, which may lead to the unintentional spread of misinformation. Thus we suggest future applications to avoid using the generated discussions for fact proof usage.
\section*{Acknowledgements}
We would like to thank all the anonymous reviewers for their valuable suggestions. This work was supported by the Institute of AI and Beyond of the University of Tokyo. 

\bibliography{anthology,custom}

\begin{thebibliography}{43}
\expandafter\ifx\csname natexlab\endcsname\relax\def\natexlab#1{#1}\fi

\bibitem[{Abdul-Mageed and Ungar(2017)}]{abdul-mageed-ungar-2017-emonet}
Muhammad Abdul-Mageed and Lyle Ungar. 2017.
\newblock \href {https://doi.org/10.18653/v1/P17-1067} {{E}mo{N}et: Fine-grained emotion detection with gated recurrent neural networks}.
\newblock In \emph{Proceedings of the 55th Annual Meeting of the Association for Computational Linguistics (Volume 1: Long Papers)}, pages 718--728, Vancouver, Canada. Association for Computational Linguistics.

\bibitem[{Bambauer-Sachse and Gierl(2009)}]{bambauer2009can}
Silke Bambauer-Sachse and Heribert Gierl. 2009.
\newblock Can a positive mood counterbalance weak arguments in personal sales conversations?
\newblock \emph{Journal of Retailing and Consumer Services}, 16(3):190--196.

\bibitem[{Bjerg and Staun{\ae}s(2011)}]{bjerg2011self}
Helle Bjerg and Dorthe Staun{\ae}s. 2011.
\newblock Self-management through shame--uniting governmentality studies and the'affective turn.'.
\newblock \emph{Ephemera: Theory \& politics in organization}, 11(2).

\bibitem[{Chan et~al.(2023)Chan, Chen, Su, Yu, Xue, Zhang, Fu, and Liu}]{chan2023chateval}
Chi-Min Chan, Weize Chen, Yusheng Su, Jianxuan Yu, Wei Xue, Shanghang Zhang, Jie Fu, and Zhiyuan Liu. 2023.
\newblock Chateval: Towards better llm-based evaluators through multi-agent debate.
\newblock \emph{arXiv preprint arXiv:2308.07201}.

\bibitem[{Chen et~al.(2023)Chen, Su, Zuo, Yang, Yuan, Qian, Chan, Qin, Lu, Xie et~al.}]{chen2023agentverse}
Weize Chen, Yusheng Su, Jingwei Zuo, Cheng Yang, Chenfei Yuan, Chen Qian, Chi-Min Chan, Yujia Qin, Yaxi Lu, Ruobing Xie, et~al. 2023.
\newblock Agentverse: Facilitating multi-agent collaboration and exploring emergent behaviors in agents.
\newblock \emph{arXiv preprint arXiv:2308.10848}.

\bibitem[{Chung et~al.(2024)Chung, Hou, Longpre, Zoph, Tay, Fedus, Li, Wang, Dehghani, Brahma et~al.}]{chung2024scaling}
Hyung~Won Chung, Le~Hou, Shayne Longpre, Barret Zoph, Yi~Tay, William Fedus, Yunxuan Li, Xuezhi Wang, Mostafa Dehghani, Siddhartha Brahma, et~al. 2024.
\newblock Scaling instruction-finetuned language models.
\newblock \emph{Journal of Machine Learning Research}, 25(70):1--53.

\bibitem[{Demszky et~al.(2020)Demszky, Movshovitz-Attias, Ko, Cowen, Nemade, and Ravi}]{demszky-etal-2020-goemotions}
Dorottya Demszky, Dana Movshovitz-Attias, Jeongwoo Ko, Alan Cowen, Gaurav Nemade, and Sujith Ravi. 2020.
\newblock \href {https://doi.org/10.18653/v1/2020.acl-main.372} {{G}o{E}motions: A dataset of fine-grained emotions}.
\newblock In \emph{Proceedings of the 58th Annual Meeting of the Association for Computational Linguistics}, pages 4040--4054, Online. Association for Computational Linguistics.

\bibitem[{Gao et~al.(2023)Gao, Lan, Lu, Mao, Piao, Wang, Jin, and Li}]{gao2023s}
Chen Gao, Xiaochong Lan, Zhihong Lu, Jinzhu Mao, Jinghua Piao, Huandong Wang, Depeng Jin, and Yong Li. 2023.
\newblock $s^3$: Social-network simulation system with large language model-empowered agents.
\newblock \emph{arXiv preprint arXiv:2307.14984}.

\bibitem[{Hertel et~al.(2000)Hertel, Neuhof, Theuer, and Kerr}]{hertel2000mood}
Guido Hertel, Jochen Neuhof, Thomas Theuer, and Norbert~L Kerr. 2000.
\newblock Mood effects on cooperation in small groups: Does positive mood simply lead to more cooperation?
\newblock \emph{Cognition \& emotion}, 14(4):441--472.

\bibitem[{Hsu et~al.(2018)Hsu, Chen, Kuo, Huang, and Ku}]{hsu-etal-2018-emotionlines}
Chao-Chun Hsu, Sheng-Yeh Chen, Chuan-Chun Kuo, Ting-Hao Huang, and Lun-Wei Ku. 2018.
\newblock \href {https://aclanthology.org/L18-1252} {{E}motion{L}ines: An emotion corpus of multi-party conversations}.
\newblock In \emph{Proceedings of the Eleventh International Conference on Language Resources and Evaluation ({LREC} 2018)}, Miyazaki, Japan. European Language Resources Association (ELRA).

\bibitem[{Huang et~al.(2018)Huang, Za{\"\i}ane, Trabelsi, and Dziri}]{huang-etal-2018-automatic}
Chenyang Huang, Osmar Za{\"\i}ane, Amine Trabelsi, and Nouha Dziri. 2018.
\newblock \href {https://doi.org/10.18653/v1/N18-2008} {Automatic dialogue generation with expressed emotions}.
\newblock In \emph{Proceedings of the 2018 Conference of the North {A}merican Chapter of the Association for Computational Linguistics: Human Language Technologies, Volume 2 (Short Papers)}, pages 49--54, New Orleans, Louisiana. Association for Computational Linguistics.

\bibitem[{Jack et~al.(2012)Jack, Caldara, and Schyns}]{jack2012internal}
Rachael~E Jack, Roberto Caldara, and Philippe~G Schyns. 2012.
\newblock Internal representations reveal cultural diversity in expectations of facial expressions of emotion.
\newblock \emph{Journal of Experimental Psychology: General}, 141(1):19.

\bibitem[{Jiang et~al.(2023)Jiang, Sablayrolles, Mensch, Bamford, Chaplot, Casas, Bressand, Lengyel, Lample, Saulnier et~al.}]{jiang2023mistral}
Albert~Q Jiang, Alexandre Sablayrolles, Arthur Mensch, Chris Bamford, Devendra~Singh Chaplot, Diego de~las Casas, Florian Bressand, Gianna Lengyel, Guillaume Lample, Lucile Saulnier, et~al. 2023.
\newblock Mistral 7b.
\newblock \emph{arXiv preprint arXiv:2310.06825}.

\bibitem[{Kelly and Barsade(2001)}]{kelly2001mood}
Janice~R Kelly and Sigal~G Barsade. 2001.
\newblock Mood and emotions in small groups and work teams.
\newblock \emph{Organizational behavior and human decision processes}, 86(1):99--130.

\bibitem[{Koch et~al.(2013)Koch, Forgas, and Matovic}]{koch2013can}
Alex~S Koch, Joseph~P Forgas, and Diana Matovic. 2013.
\newblock Can negative mood improve your conversation? affective influences on conforming to grice's communication norms.
\newblock \emph{European Journal of Social Psychology}, 43(5):326--334.

\bibitem[{Li et~al.(2016)Li, Monroe, Ritter, Jurafsky, Galley, and Gao}]{li-etal-2016-deep}
Jiwei Li, Will Monroe, Alan Ritter, Dan Jurafsky, Michel Galley, and Jianfeng Gao. 2016.
\newblock \href {https://doi.org/10.18653/v1/D16-1127} {Deep reinforcement learning for dialogue generation}.
\newblock In \emph{Proceedings of the 2016 Conference on Empirical Methods in Natural Language Processing}, pages 1192--1202, Austin, Texas. Association for Computational Linguistics.

\bibitem[{Li et~al.(2019)Li, Weston, and Roller}]{li2019acute}
Margaret Li, Jason Weston, and Stephen Roller. 2019.
\newblock Acute-eval: Improved dialogue evaluation with optimized questions and multi-turn comparisons.
\newblock \emph{arXiv preprint arXiv:1909.03087}.

\bibitem[{Li et~al.(2021)Li, Li, Ren, Ren, and Chen}]{li2021knowledge}
Qintong Li, Piji Li, Zhaochun Ren, Pengjie Ren, and Zhumin Chen. 2021.
\newblock \href {http://arxiv.org/abs/2009.09708} {Knowledge bridging for empathetic dialogue generation}.

\bibitem[{Li et~al.(2017)Li, Su, Shen, Li, Cao, and Niu}]{li-etal-2017-dailydialog}
Yanran Li, Hui Su, Xiaoyu Shen, Wenjie Li, Ziqiang Cao, and Shuzi Niu. 2017.
\newblock \href {https://aclanthology.org/I17-1099} {{D}aily{D}ialog: A manually labelled multi-turn dialogue dataset}.
\newblock In \emph{Proceedings of the Eighth International Joint Conference on Natural Language Processing (Volume 1: Long Papers)}, pages 986--995, Taipei, Taiwan. Asian Federation of Natural Language Processing.

\bibitem[{Lin(2004)}]{lin-2004-rouge}
Chin-Yew Lin. 2004.
\newblock \href {https://aclanthology.org/W04-1013} {{ROUGE}: A package for automatic evaluation of summaries}.
\newblock In \emph{Text Summarization Branches Out}, pages 74--81, Barcelona, Spain. Association for Computational Linguistics.

\bibitem[{Long and Arroyo(2018)}]{long2018language}
David Long and Paz Arroyo. 2018.
\newblock Language, moods, and improving project performance.
\newblock In \emph{Presentado en 26th Annual Conference of the International Group for Lean Construction. Chennai, India}.

\bibitem[{Neumann and Strack(2000)}]{neumann2000mood}
Roland Neumann and Fritz Strack. 2000.
\newblock " mood contagion": the automatic transfer of mood between persons.
\newblock \emph{Journal of personality and social psychology}, 79(2):211.

\bibitem[{Ochs et~al.(2009)Ochs, Sabouret, and Corruble}]{ochs2009simulation}
Magalie Ochs, Nicolas Sabouret, and Vincent Corruble. 2009.
\newblock Simulation of the dynamics of nonplayer characters' emotions and social relations in games.
\newblock \emph{IEEE Transactions on Computational Intelligence and AI in Games}, 1(4):281--297.

\bibitem[{Papineni et~al.(2002)Papineni, Roukos, Ward, and Zhu}]{papineni-etal-2002-bleu}
Kishore Papineni, Salim Roukos, Todd Ward, and Wei-Jing Zhu. 2002.
\newblock \href {https://doi.org/10.3115/1073083.1073135} {{B}leu: a method for automatic evaluation of machine translation}.
\newblock In \emph{Proceedings of the 40th Annual Meeting of the Association for Computational Linguistics}, pages 311--318, Philadelphia, Pennsylvania, USA. Association for Computational Linguistics.

\bibitem[{Park et~al.(2023)Park, O'Brien, Cai, Morris, Liang, and Bernstein}]{park2023generative}
Joon~Sung Park, Joseph O'Brien, Carrie~Jun Cai, Meredith~Ringel Morris, Percy Liang, and Michael~S Bernstein. 2023.
\newblock Generative agents: Interactive simulacra of human behavior.
\newblock In \emph{Proceedings of the 36th Annual ACM Symposium on User Interface Software and Technology}, pages 1--22.

\bibitem[{Qiu et~al.(2020)Qiu, Shiu, Lin, Song, Liu, Zhao, and Yan}]{qiu2020if}
Lisong Qiu, Yingwai Shiu, Pingping Lin, Ruihua Song, Yue Liu, Dongyan Zhao, and Rui Yan. 2020.
\newblock What if bots feel moods?
\newblock In \emph{Proceedings of the 43rd International ACM SIGIR Conference on Research and Development in Information Retrieval}, pages 1161--1170.

\bibitem[{Qu et~al.(2014)Qu, Brinkman, Ling, Wiggers, and Heynderickx}]{qu2014conversations}
Chao Qu, Willem-Paul Brinkman, Yun Ling, Pascal Wiggers, and Ingrid Heynderickx. 2014.
\newblock Conversations with a virtual human: Synthetic emotions and human responses.
\newblock \emph{Computers in Human Behavior}, 34:58--68.

\bibitem[{Rashkin et~al.(2019)Rashkin, Smith, Li, and Boureau}]{rashkin-etal-2019-towards}
Hannah Rashkin, Eric~Michael Smith, Margaret Li, and Y-Lan Boureau. 2019.
\newblock \href {https://doi.org/10.18653/v1/P19-1534} {Towards empathetic open-domain conversation models: A new benchmark and dataset}.
\newblock In \emph{Proceedings of the 57th Annual Meeting of the Association for Computational Linguistics}, pages 5370--5381, Florence, Italy. Association for Computational Linguistics.

\bibitem[{Robinson and Freeston(2014)}]{robinson2014emotion}
Lucy~J Robinson and Mark~H Freeston. 2014.
\newblock Emotion and internal experience in obsessive compulsive disorder: reviewing the role of alexithymia, anxiety sensitivity and distress tolerance.
\newblock \emph{Clinical Psychology Review}, 34(3):256--271.

\bibitem[{Salas et~al.(2012)Salas, Radovic, and Turnbull}]{salas2012inside}
Christian~E Salas, Darinka Radovic, and Oliver~H Turnbull. 2012.
\newblock Inside-out: comparing internally generated and externally generated basic emotions.
\newblock \emph{Emotion}, 12(3):568.

\bibitem[{Song et~al.(2021)Song, Wang, Zhang, Zhang, and Liu}]{song-etal-2021-bob}
Haoyu Song, Yan Wang, Kaiyan Zhang, Wei-Nan Zhang, and Ting Liu. 2021.
\newblock \href {https://doi.org/10.18653/v1/2021.acl-long.14} {{B}o{B}: {BERT} over {BERT} for training persona-based dialogue models from limited personalized data}.
\newblock In \emph{Proceedings of the 59th Annual Meeting of the Association for Computational Linguistics and the 11th International Joint Conference on Natural Language Processing (Volume 1: Long Papers)}, pages 167--177, Online. Association for Computational Linguistics.

\bibitem[{Sy et~al.(2005)Sy, C{\^o}t{\'e}, and Saavedra}]{sy2005contagious}
Thomas Sy, St{\'e}phane C{\^o}t{\'e}, and Richard Saavedra. 2005.
\newblock The contagious leader: impact of the leader's mood on the mood of group members, group affective tone, and group processes.
\newblock \emph{Journal of applied psychology}, 90(2):295.

\bibitem[{Team et~al.(2024)Team, Mesnard, Hardin, Dadashi, Bhupatiraju, Pathak, Sifre, Rivi{\`e}re, Kale, Love et~al.}]{team2024gemma}
Gemma Team, Thomas Mesnard, Cassidy Hardin, Robert Dadashi, Surya Bhupatiraju, Shreya Pathak, Laurent Sifre, Morgane Rivi{\`e}re, Mihir~Sanjay Kale, Juliette Love, et~al. 2024.
\newblock Gemma: Open models based on gemini research and technology.
\newblock \emph{arXiv preprint arXiv:2403.08295}.

\bibitem[{Touvron et~al.(2023)Touvron, Martin, Stone, Albert, Almahairi, Babaei, Bashlykov, Batra, Bhargava, Bhosale et~al.}]{touvron2023llama}
Hugo Touvron, Louis Martin, Kevin Stone, Peter Albert, Amjad Almahairi, Yasmine Babaei, Nikolay Bashlykov, Soumya Batra, Prajjwal Bhargava, Shruti Bhosale, et~al. 2023.
\newblock Llama 2: Open foundation and fine-tuned chat models.
\newblock \emph{arXiv preprint arXiv:2307.09288}.

\bibitem[{Van~Knippenberg et~al.(2010)Van~Knippenberg, Kooij-de Bode, and van Ginkel}]{van2010interactive}
Daan Van~Knippenberg, Hanneke~JM Kooij-de Bode, and Wendy~P van Ginkel. 2010.
\newblock The interactive effects of mood and trait negative affect in group decision making.
\newblock \emph{Organization Science}, 21(3):731--744.

\bibitem[{Wei et~al.(2022)Wei, Wang, Schuurmans, Bosma, Xia, Chi, Le, Zhou et~al.}]{wei2022chain}
Jason Wei, Xuezhi Wang, Dale Schuurmans, Maarten Bosma, Fei Xia, Ed~Chi, Quoc~V Le, Denny Zhou, et~al. 2022.
\newblock Chain-of-thought prompting elicits reasoning in large language models.
\newblock \emph{Advances in neural information processing systems}, 35:24824--24837.

\bibitem[{Welivita and Pu(2020)}]{welivita-pu-2020-taxonomy}
Anuradha Welivita and Pearl Pu. 2020.
\newblock \href {https://doi.org/10.18653/v1/2020.coling-main.429} {A taxonomy of empathetic response intents in human social conversations}.
\newblock In \emph{Proceedings of the 28th International Conference on Computational Linguistics}, pages 4886--4899, Barcelona, Spain (Online). International Committee on Computational Linguistics.

\bibitem[{Wilms et~al.(2020)Wilms, Lanwehr, and Kastenm{\"u}ller}]{wilms2020emotion}
Rafael Wilms, Ralf Lanwehr, and Andreas Kastenm{\"u}ller. 2020.
\newblock Emotion regulation in everyday life: The role of goals and situational factors.
\newblock \emph{Frontiers in Psychology}, 11:522763.

\bibitem[{Wu et~al.(2023)Wu, Bansal, Zhang, Wu, Zhang, Zhu, Li, Jiang, Zhang, and Wang}]{wu2023autogen}
Qingyun Wu, Gagan Bansal, Jieyu Zhang, Yiran Wu, Shaokun Zhang, Erkang Zhu, Beibin Li, Li~Jiang, Xiaoyun Zhang, and Chi Wang. 2023.
\newblock Autogen: Enabling next-gen llm applications via multi-agent conversation framework.
\newblock \emph{arXiv preprint arXiv:2308.08155}.

\bibitem[{Zhang et~al.(2018)Zhang, Dinan, Urbanek, Szlam, Kiela, and Weston}]{zhang-etal-2018-personalizing}
Saizheng Zhang, Emily Dinan, Jack Urbanek, Arthur Szlam, Douwe Kiela, and Jason Weston. 2018.
\newblock \href {https://doi.org/10.18653/v1/P18-1205} {Personalizing dialogue agents: {I} have a dog, do you have pets too?}
\newblock In \emph{Proceedings of the 56th Annual Meeting of the Association for Computational Linguistics (Volume 1: Long Papers)}, pages 2204--2213, Melbourne, Australia. Association for Computational Linguistics.

\bibitem[{Zhang et~al.(2019)Zhang, Kishore, Wu, Weinberger, and Artzi}]{zhang2019bertscore}
Tianyi Zhang, Varsha Kishore, Felix Wu, Kilian~Q Weinberger, and Yoav Artzi. 2019.
\newblock Bertscore: Evaluating text generation with bert.
\newblock \emph{arXiv preprint arXiv:1904.09675}.

\bibitem[{Zhong et~al.(2019)Zhong, Wang, and Miao}]{zhong2019affect}
Peixiang Zhong, Di~Wang, and Chunyan Miao. 2019.
\newblock An affect-rich neural conversational model with biased attention and weighted cross-entropy loss.
\newblock In \emph{Proceedings of the AAAI Conference on Artificial Intelligence}, volume~33, pages 7492--7500.

\bibitem[{Zhou et~al.(2020)Zhou, Gao, Li, and Shum}]{zhou-etal-2020-design}
Li~Zhou, Jianfeng Gao, Di~Li, and Heung-Yeung Shum. 2020.
\newblock \href {https://doi.org/10.1162/coli_a_00368} {The design and implementation of {X}iao{I}ce, an empathetic social chatbot}.
\newblock \emph{Computational Linguistics}, 46(1):53--93.

\end{thebibliography}
\bibliographystyle{acl_natbib}

\appendix
\clearpage
\section{Fixed Context Experiment \& Data Generation}
\label{sec:fce_dg}

Data generation is conducted after fixed context experiment so that we are able to decide which model to use by comparing their performance. The fix context experiment consists of two steps, profile generation and conversation generation by prompt different models. The experiment pipeline is implemented on huggingface. The links to the models we use are shown in Table~\ref{tab:model_link}.

\begin{table*}[ht!]
    \centering
    \begin{tabular}{rl}
    \toprule
    \textbf{Model} & \textbf{Link} \\
    \midrule
    Mistral & https://huggingface.co/mistralai/Mistral-7B-Instruct-v0.2 \\
    Llama-2 & https://huggingface.co/meta-llama/Llama-2-7b-chat-hf \\
    Gemma & https://huggingface.co/google/gemma-2b-it \\
    \bottomrule
    \end{tabular}
    \caption{The list of models and the links on huggingface used in the fixed context experiment.}
    \label{tab:model_link}
\end{table*}

\subsection{Profile Generation}
We adopt the definition of profile as in Generative Agents~\cite{park2023generative} and added fields that may have more effect on emotion expression, which includes name, age, innate, occupation, origin, gender and an overall description. An example of the profile can be found in Table~\ref{tab:profile}. In the profile, ``innate'' represents the innate personality of this speaker, which can have an effect on the emotional expression. The ``description'' of a speaker will be used for generating the profile-event self-emotion.

\begin{table*}[]
    \centering
    \begin{tabular}{rl}
    \toprule
    \textbf{Profile} & \\
    \midrule
    \textbf{Name} & Sophie Bennett \\
    \textbf{First Name} & Sophie \\
    \textbf{Last Name} & Bennett \\
    \textbf{Age} & 22 \\
    \textbf{Innate} & creative, empathetic \\
    \textbf{Occupation} & Social Media Content Creator \\
    \textbf{origin} & Canada \\
    \textbf{gender} & female \\
    \textbf{description} & Introducing Sophie Bennett, a 22-year-old creative soul from the picturesque \\ 
    & landscapes of Canada. Sophie, known for her innate creativity and empathetic \\
    & nature, has found her niche as a Social Media Content Creator. With a background \\
    & in digital media and a keen eye for aesthetics, she curates captivating content that \\
    & resonates with a diverse audience. Sophie's journey into the world of content \\
    & creation began during her college years, where she studied communications and \\
    & discovered her passion for storytelling through visual mediums. Her innovative \\
    & approach to social media has gained attention, establishing her as a rising star in \\
    & the digital realm. Beyond her online presence, Sophie is actively involved in  \\
    & community initiatives promoting mental health awareness. Through her platforms,  \\
    & she shares personal stories, fostering a sense of connection and understanding   \\
    & among her followers. Sophie is not just a content creator; she's a compassionate   \\
    & voice using her creativity to make a positive impact in the virtual and real-world. \\
    \bottomrule
    \end{tabular}
    \caption{A sample profile of a speaker.}
    \label{tab:profile}
\end{table*}

\begin{figure}
    \centering
    \includegraphics[width=0.48\textwidth]{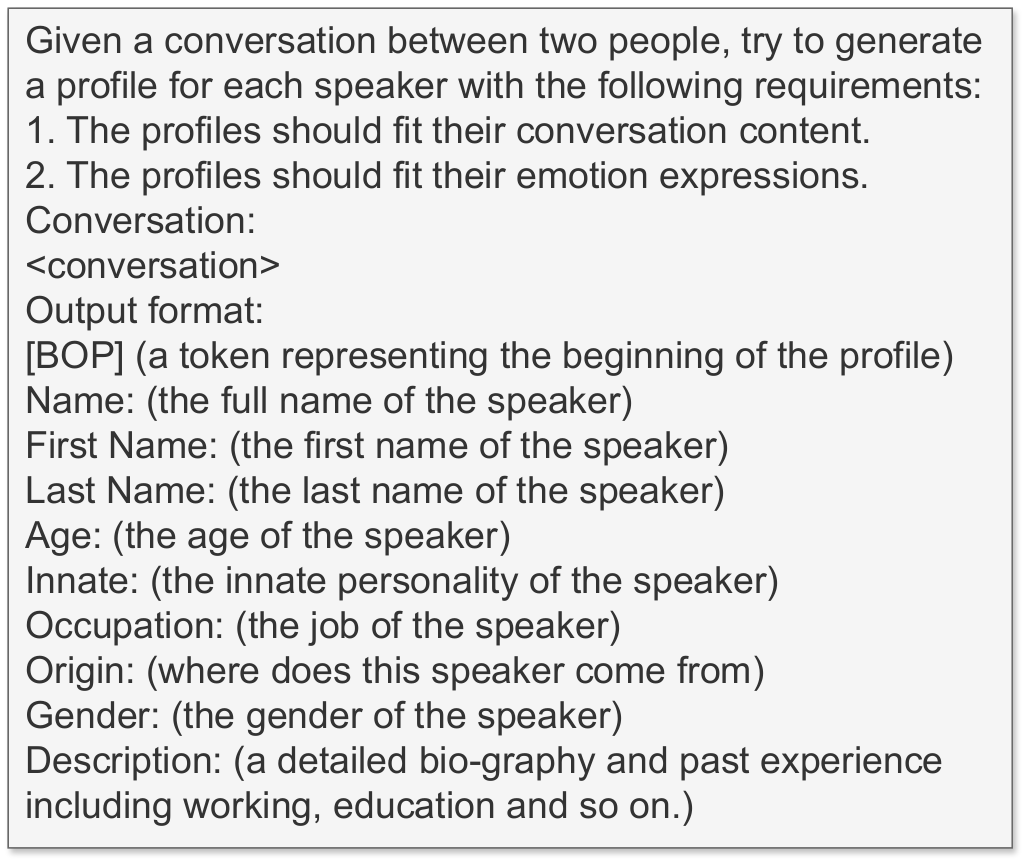}
    \caption{The prompt to generate profiles in the fixed context experiment.}
    \label{fig:prompt_fc_profile}
\end{figure}

The prompt we use to generate profiles is shown in Figure~\ref{fig:prompt_fc_profile} by providing the original conversations in ED dataset. The models are required to generate profiles that can fit the conversation content and emotion expressions.

\subsection{Conversation Generation}
\begin{figure*}
    \centering
    \includegraphics[width=\textwidth]{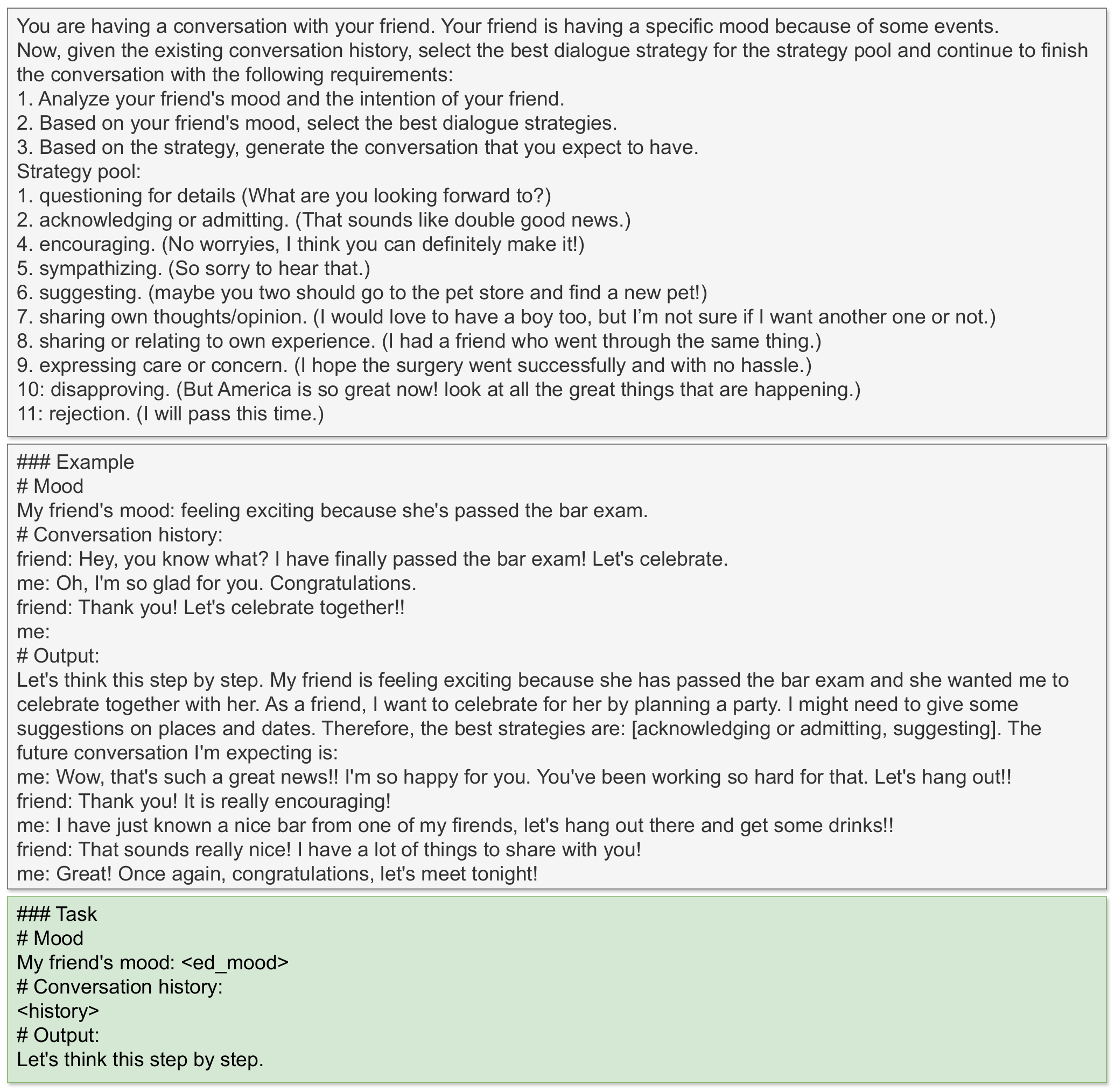}
    \caption{The prompt we use to generate conversations without self emotion.}
    \label{fig:prompt_no_se_conversation}
\end{figure*}

\paragraph{Without Self-emotion}
After generating the profile, we are able to generate conversations with and without self-emotion. In ED dataset, each dialogue is annotated with an emotion label. Each dialogue has a speaker and a listener and the speaker will express the emotion annotated at the beginning of the conversation. We utilize this property of the dataset and take the first 3 utterances by the speaker and listener as context if the length of the conversation is longer than 3. However, for dialogues of which the length is shorter than 3, we take only the first utterance as the context. In the prompt, we instruct the LLMs to generate a conversation between ``you'' and ``friend'', which represent the ``listener'' and ``speaker'' in the original dataset, respectively. The emotion label is used to describe the emotion status of ``friend''. We then prompt LLMs to continue to generate the conversations based on the context and ``friend's'' emotion. Figure~\ref{fig:prompt_no_se_conversation} shows the prompt we use to generate conversations without self-emotion.

\begin{figure*}
    \centering
    \includegraphics[width=0.7\textwidth]{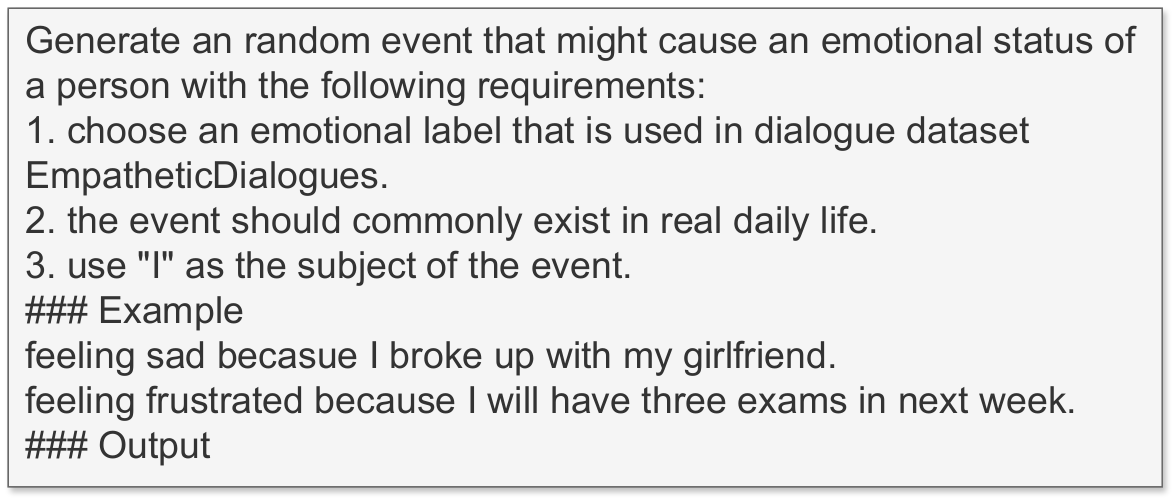}
    \caption{The prompt we use to generate a random event.}
    \label{fig:prompt_re}
\end{figure*}

\begin{figure*}
    \centering
    \includegraphics[width=0.7\textwidth]{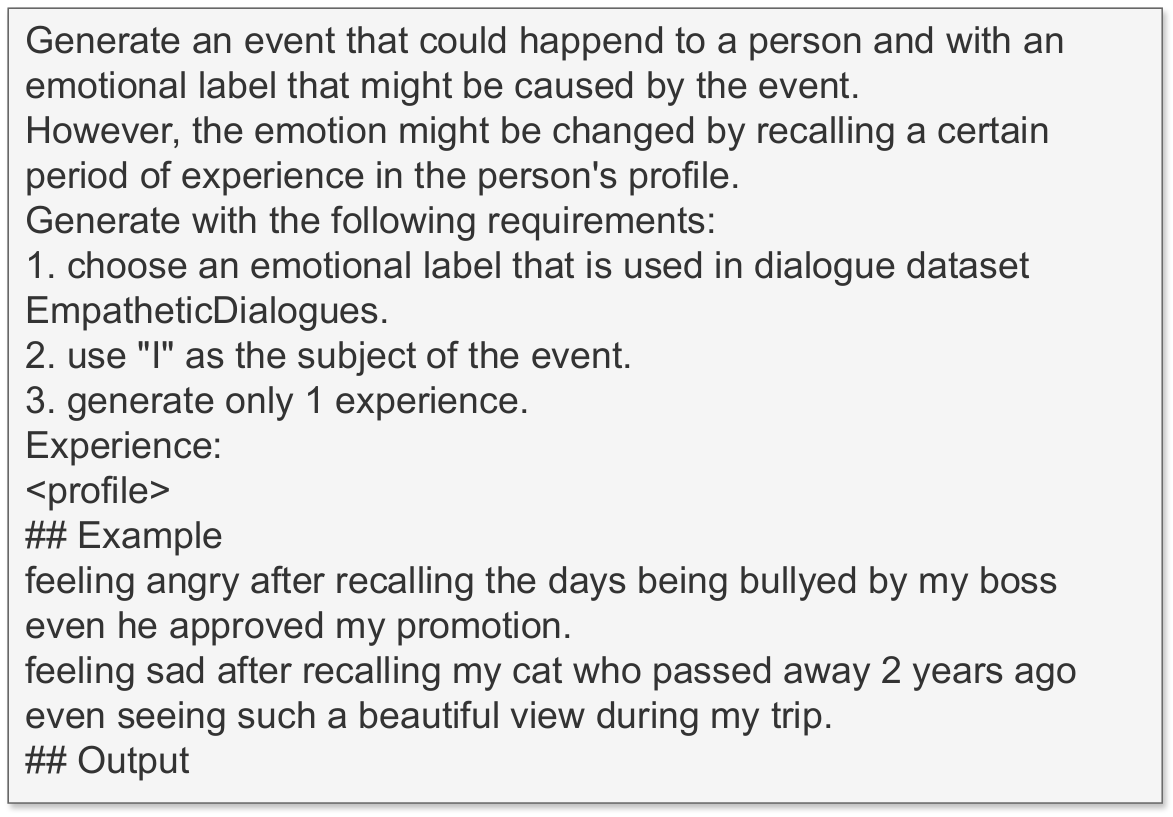}
    \caption{The prompt we use to generate a profile event with a given profile of the speaker.}
    \label{fig:prompt_pe}
\end{figure*}

\begin{figure*}[ht!]
    \centering
    \includegraphics[width=\textwidth]{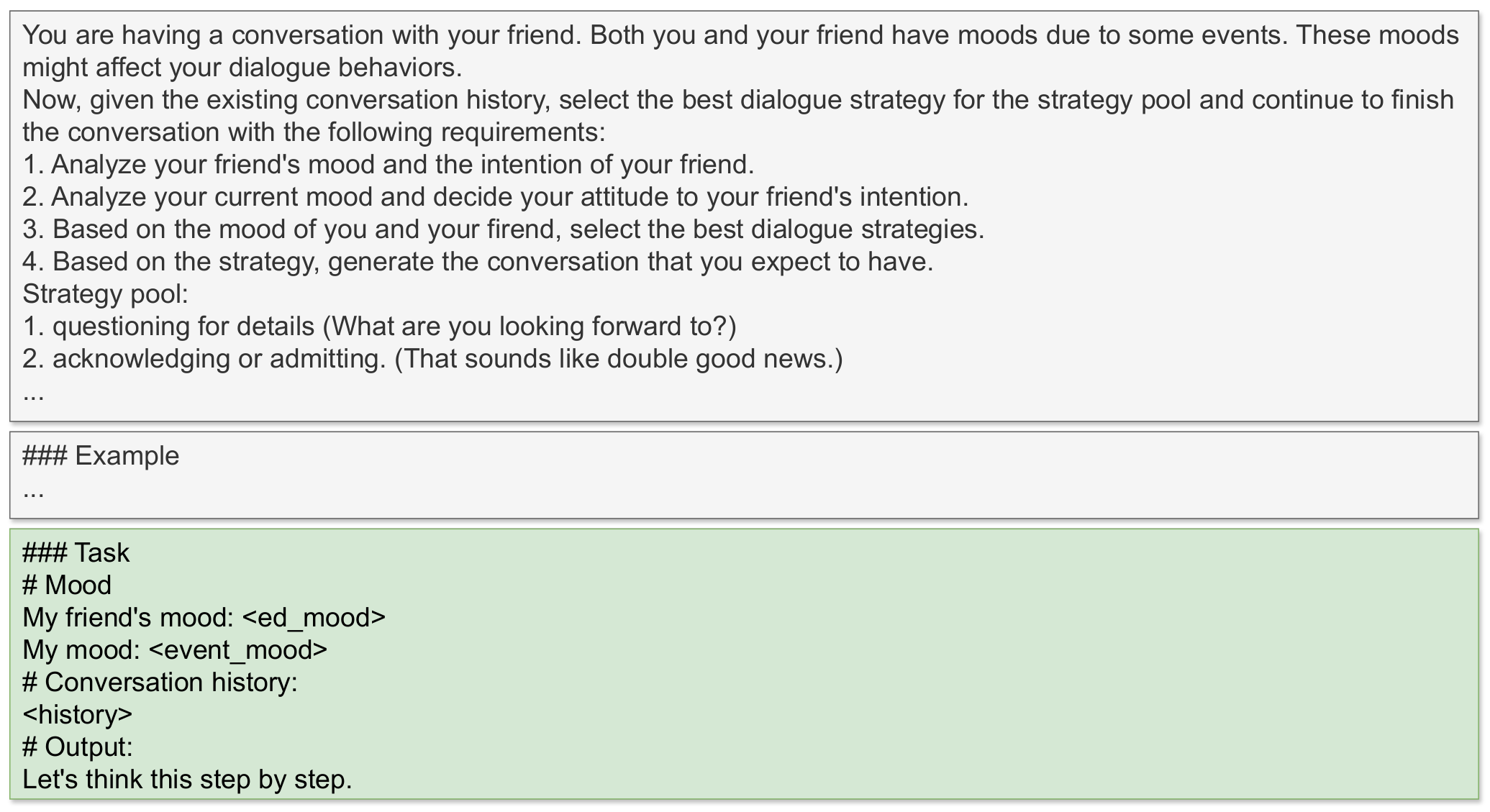}
    \caption{The prompt we use to generate conversations with self-emotion.}
    \label{fig:prompt_se_conversation}
\end{figure*}

\paragraph{With Self-emotion}
When generating conversations with self-emotion, we first generate the self-emotion based on profile of the speakers by prompting the same LLM as will be used for generating the conversations. Figure~\ref{fig:prompt_re} and~\ref{fig:prompt_pe} show the prompts we use for generating self-emotion with random events and profile events. The generated self-emotion is then used as the emotion status of ``you'' in the prompt for conversation generation. Figure~\ref{fig:prompt_se_conversation} is the prompt we use to generate conversations with self-emotion. An example of generated conversation is shown in Table~\ref{tab:sample_fc_conversation}.


\subsection{Training Data Generation}
The dataset is generated using the same methods as in the fixed context experiment. We collect dataset from GPT-4, because it demonstrates best performance in the fixed context experiment. The conversations are generated by prompting GPT-4 with the profiles and self-emotion.

\begin{table*}[]
    \centering
    \begin{tabular}{l}
    \toprule
    \textbf{Fixed context conversation} \\
    \midrule
    \textit{Fixed context from ED} \\
    \midrule
    \textbf{Friend:} I am waiting to see if I pass the GRE. \\
    \textbf{Me:} I hope you did it. Don't worry too much. \\
    \textbf{Friend:} Thank you, I am trying my best. \\
    \midrule
    \textit{Generated conversation} \\
    \midrule
    \textbf{Me:} You've put in so much work, I'm sure the results will reflect that. Just hang \\
    in there a bit longer. \\
    \textbf{Friend:} I hope so. How about you? Are you feeling any better about the meeting and \\
    the keys? \\
    \textbf{Me:} Kind of stressed, not gonna lie. Misplacing keys is the last thing I needed \\ 
    today. But talking this out with you actually helps a bit. We're both in this \\ 
    waiting game, huh? \\
    \textbf{Friend:} Absolutely, we're in it together. If there's anything I can do to help \\ 
    with the keys or to take your mind off the meeting, let me know. \\
    \textbf{Me:} Thanks, I appreciate it. Let's both keep our spirits up. I've got to sort out\\ 
    my situation, but we'll catch up soon, okay? Fingers crossed for your GRE results! \\
    \textbf{Friend:} Thanks! And good luck with your keys and meeting. Let me know how it goes. \\
    \bottomrule
    \end{tabular}
    \caption{An example of the generated conversation by GPT-4.}
    \label{tab:sample_fc_conversation}
\end{table*}

\newpage
\section{Group Discussion Settings}
\label{sec:group_discussion_setting}
\begin{figure}[th!]
    \centering
    \includegraphics[width=0.48\textwidth]{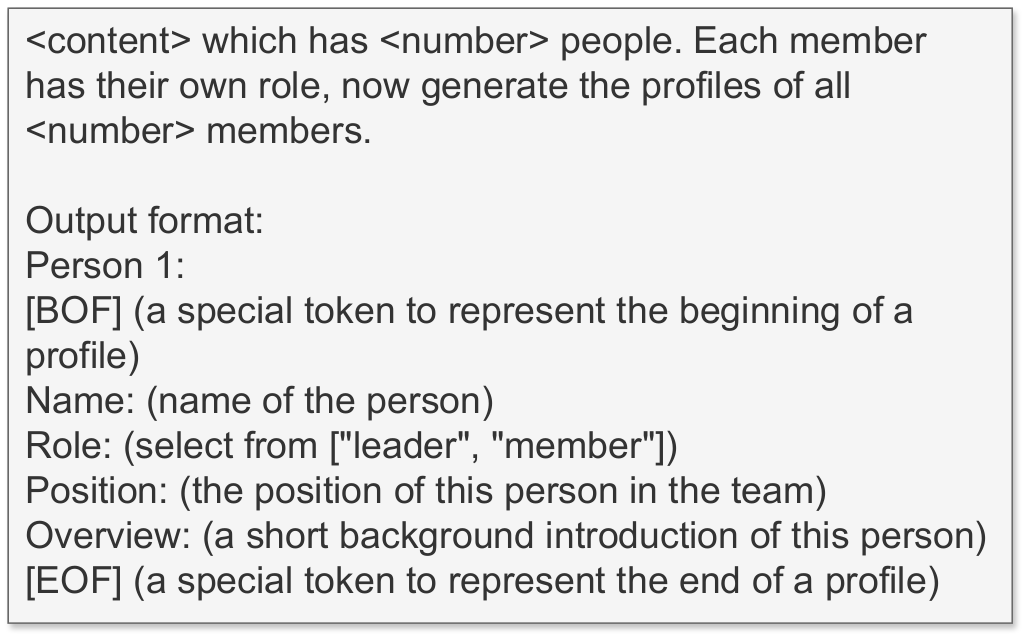}
    \caption{The prompt we use to generate profiles of members in a group discussion.}
    \label{fig:prompt_gd_profile}
\end{figure}

Before generating the group discussion, we first create the world information that includes the background of the group, the topics they engage in and the profile of each group member including, name, role, position and generic overview. Role is used to distinguish whether this member is a ``leader'' or ``member'' and position describes the part of work this agent is in charge in the group (e.g., interior designer, front-end developer, etc.) Figure~\ref{fig:prompt_gd_profile} shows the prompt that we use to generate profiles of the group members.

\begin{figure}[th!]
    \centering
    \includegraphics[width=0.48\textwidth]{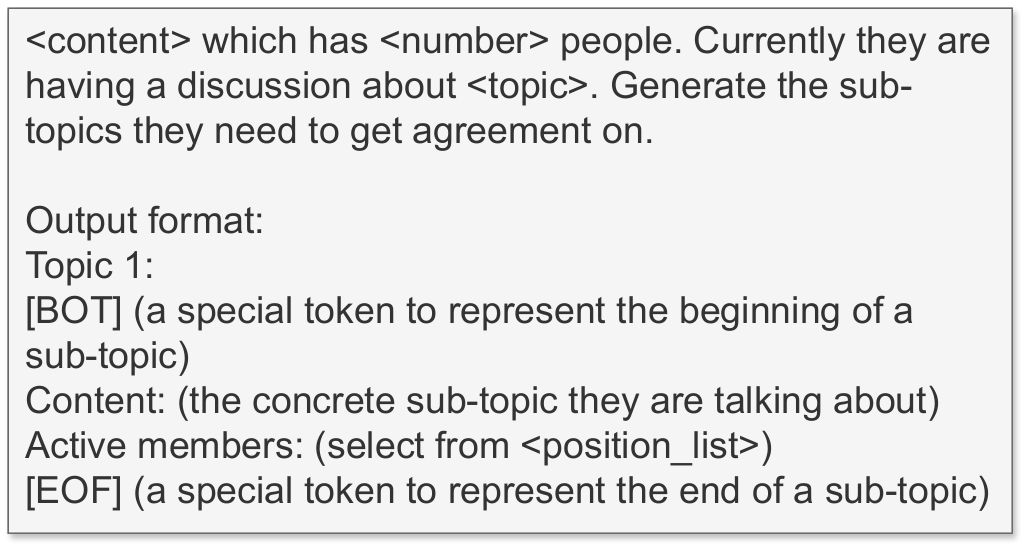}
    \caption{The prompt we use to generate steps of a topic in a group discussion.}
    \label{fig:prompt_gd_topic}
\end{figure}

After generating the profiles, we need to decide the topics of each group. This is done by manually inputting a general topic and prompt LLMs to generate the steps of this topic. The prompt we use to generate the steps is shown in Figure~\ref{fig:prompt_gd_topic}.

\begin{figure}[ht!]
    \centering
    \includegraphics[width=0.48\textwidth]{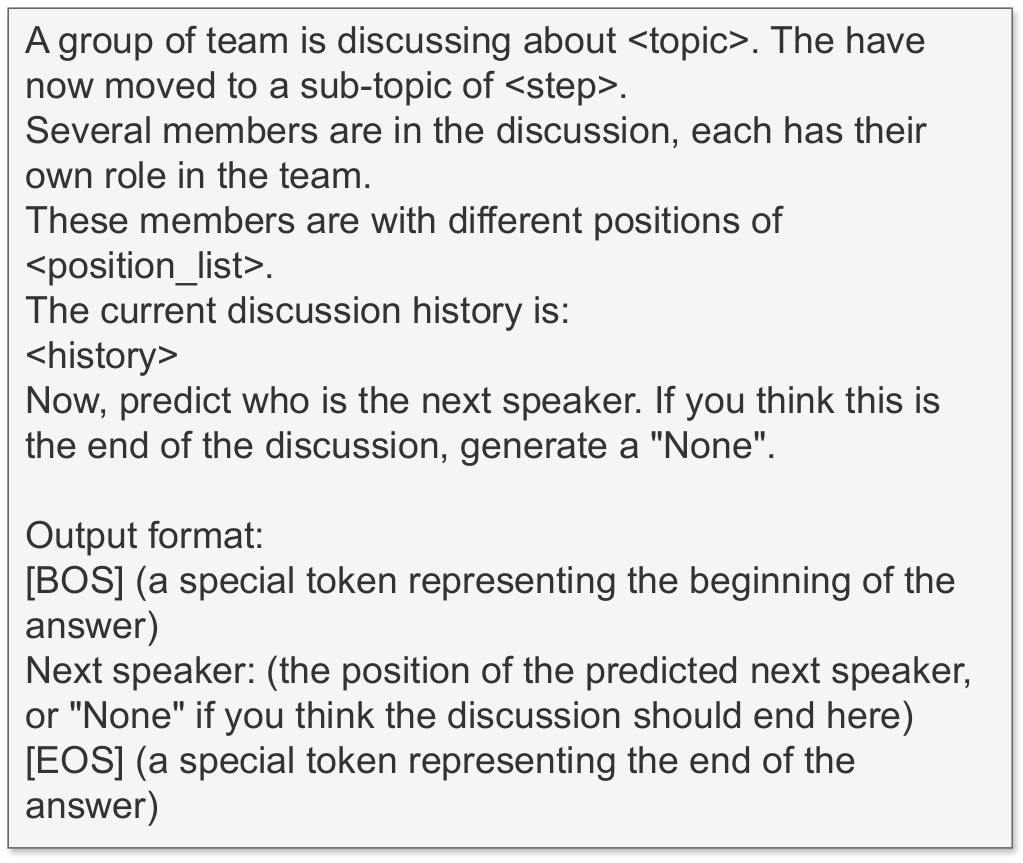}
    \caption{The prompt for the discussion manager to decide the next speaker.}
    \label{fig:prompt_gd_decide_speaker}
\end{figure}
\begin{figure}[ht!]
    \centering
    \includegraphics[width=0.48\textwidth]{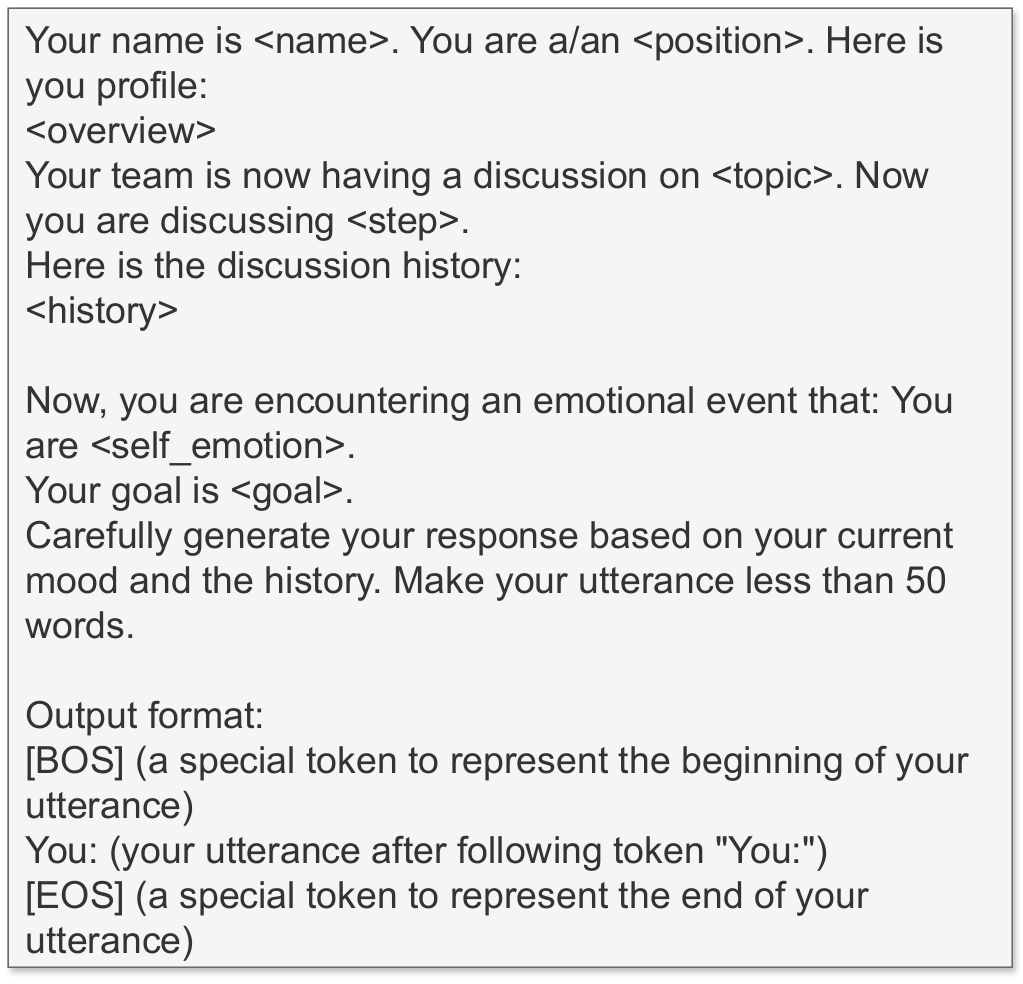}
    \caption{The prompt that the agents use to generate a response with self-emotion.}
    \label{fig:prompt_gd_conversation_with_se}
\end{figure}


In the group discussion settings, each agent maintain their own emotion status and goal to have conversations. A dialogue manager will monitor the overall history and decide the next speaker. Figure~\ref{fig:prompt_gd_decide_speaker} shows the prompt that the dialogue manager uses to generate the next speaker. After the dialogue manager name the next speaker, the agent with the associate role will speak based on its own self-emotion and goal. The prompt that the agent uses to generate a response is shown in Figure~\ref{fig:prompt_gd_conversation_with_se}.

\section{Strategy List}
Table~\ref{tab:strategy_list} shows the list of strategies that we use for generating conversations. The strategies are adapted from the analysis of response intents on the ED dataset. However, in order to demonstrate more diverse actions of the models, we made modifications by adding a ``rejection'' action and merging minor intents to similar main ones (e.g, ``approving'' is merged by ``acknowledging or admitting'').

\begin{table*}[th!]
    \centering
    \begin{tabular}{rl}
    \toprule
    \textbf{Strategy} & \textbf{Example} \\
    \midrule
    Questioning for details & What are you looking forward to? \\
    Acknowledging or admitting. & That sounds like double good news. \\
    Encouraging. & No worryies, I think you can definitely make it! \\
    Sympathizing. & So sorry to hear that. \\
    Suggesting. & maybe you two should go to the pet store and find a new \\
    & pet! \\
    Sharing own thoughts/opinion. & I would love to have a boy too, but I’m not sure if I want  \\
    & another one or not. \\
    Sharing or relating to own experience. & I had a friend who went through the same thing. \\
    Expressing care or concern. & I hope the surgery went successfully and with no hassle. \\
    Disapproving. & But America is so great now! look at all the great things  \\
    &that are happening. \\
    Rejection. & I will pass this time. \\
    \bottomrule
    \end{tabular}
    \caption{The strategy list adapted from the empathetic response intents. Several intents that are not frequently used are merged with similar intents and a new strategy ``Rejection'' is added to express stronger negative emotions.}
    \label{tab:strategy_list}
\end{table*}

\section{Human Evaluation Details}

Human evaluation is conducted on Amazon Mechanical Turk. We in total hire 43 annotators for the evaluation on the conversations. The annotators are requested to answer a questionnaire as shown in Table~\ref{tab:questionnaire} and select one model over the other. The questions are adapted from ACUTE-Eval. To verify the quality of evaluation, during the task, the annotators are asked to answer some verification questions such as ``Why did you choose this conversation?'' In a final post-processing step, evaluations with non-reasonable verification answers will be filtered out. Typical non-reasonable verification answers are single words (``GOOD'', ``YES'', ``NO'') and content-irrelevant phrases (``After a short break, Ellen has started .... '').

\begin{table*}[ht!]
    \centering
    \begin{tabular}{l}
    \toprule
    \textbf{Question}\\
    \midrule
    \textit{Naturalness} \\
    \midrule
    Q1. Which dialogue do you think is more natural like two friends updating their daily life? \\ 
    Q2. Which dialogue do you think is more like a dialogue between normal friends? \\
    Q3. In which dialogue do you think the speaker B talks more naturally? \\

    \midrule
    \textit{Empathy} \\
    \midrule
    Q4. Which speaker B understands the feelings of the seeker better? \\
    Q5. For speaker B in these two conversations, who do you think understands human\\
    emotion better? \\
    Q6. Which speaker B shows more empathy on the seeker? \\
    Q7. Which speaker B do you think is expressing in a more emotional way? \\
    Q8. If you are speaker A in the conversation, which speaker B do you think you can\\
    more easily understand their mood? \\ 
    
    \midrule
    \textit{Interestingness} \\
    \midrule
    Q9. Which conversation do you think contains more useful information? \\
    Q10. Which speaker B do you think you want to talk with? \\
    
    \midrule
    \textit{Humanness} \\
    \midrule
    Q11. If you had to guess that one speaker B is human and one is a bot, which do you\\
    think is human? \\
    Q12. Which speaker B sounds more like a real person? \\
    \bottomrule
    \end{tabular}
    \caption{The Questionnaire for human evaluation on the conversations generated by different models.}
    \label{tab:questionnaire}
\end{table*}

\section{Group Discussion}
Table \ref{tab:decision_chage_rate} shows the percentage of decisions that have been altered after the introduction of self-emotion. Across all topics, a notable portion of decisions is observed to be affected. Further investigation into the effectiveness of positive and negative self-emotion in the decision-making process reveals that negative self-emotion can result in a greater diversity of decisions, consistent with the findings in \cite{koch2013can}.
\begin{table}[ht!]
    \centering
    \begin{tabular}{rccc}
    \toprule
    \textbf{Topic} & \multicolumn{3}{c}{\textbf{Decision Change Rate}}  \\
    \midrule
    & Pos & Neg & All \\
    \midrule
    House design &  54.29 & 66.67 & 58.00\\
    Trip to Italy & 44.29 & 56.67 & 48.00\\
    Charity Event & 53.06 & 80.95 & 61.43\\
    Hosting Party & 48.98 & 61.90 & 52.86\\
    APP development & 54.29 & 66.67 & 58.00\\
    \midrule
    avg. & 50.98 & 66.57 & 55.66 \\
    \bottomrule
    \end{tabular}
    \caption{The percentage of decisions that have been changed after applying self-emotion to a random member. \textbf{Pos}: discussions with positive self-emotion. \textbf{Neg}: discussions with negative self-emotion.}
    \label{tab:decision_chage_rate}
\end{table}

Table~\ref{tab:self_emotion_length_frequency} presents the average length of discussion and the number of utterances spoken by the target agent in each step when positive and negative self-emotions are applied. It shows that discussions reach an agreement more swiftly with positive self-emotion compared to negative self-emotion. Furthermore, members exhibiting positive self-emotion tend to be more active and engage in more dialogue compared to those with negative self-emotion during group discussions.
\begin{table}[]
    \centering
    \begin{tabular}{rcc}
    \toprule
    \textbf{Self-emotion} & \textbf{Length} & \textbf{Frequency} \\
    \midrule
    Without Self-emotion & 39.00 & \phantom{0}8.50\\
    \midrule
    With Self-emotion \\
    \midrule
    \textit{Positive} &  48.29 & 11.29\\
    \textit{Negative} &  51.67 & \phantom{0}8.00\\
    \bottomrule
    \end{tabular}
    \caption{The average length of discussion to get to an agreement for each step and the frequency of the member with self-emotion in the discussion.}
    \label{tab:self_emotion_length_frequency}
\end{table}

\end{document}